\documentclass[graybox, secnum]{svmult}

\usepackage{mathptmx}       
\usepackage{helvet}         
\usepackage{courier}        
\usepackage{type1cm}        
\usepackage{makeidx}         
\usepackage{graphicx}        
\usepackage{multicol}        
\usepackage[bottom]{footmisc}
\usepackage{hyperref}        
\usepackage{soul}            
\hypersetup{colorlinks=true,urlcolor=blue}
\usepackage{amsmath}  
\usepackage{amssymb} 

\usepackage[square,numbers]{natbib}

\newcommand{\aap}{{A\&A}}
\newcommand{\aaps}{{A\&AS}}
\newcommand{\aapr}{{A\&A Rev.}}

\newcommand{\apj}{{ApJ}}

\newcommand{\apjl}{{ApJ}}
\newcommand{\apjs}{{ApJS}}

\newcommand{\araa}{{ARAA}}

\newcommand{\mnras}{{MNRAS}}

\newcommand{\nat}{{Nat}}

\newcommand{\pasj}{{PASJ}}

\newcommand{\prd}{{Phys. Rev. D}}

\newcommand{\kms}{{km\,s$^{-1}$}}
\renewcommand{\bar}{\overline}

\newcommand{{\ism}}{interstellar medium}
\newcommand{{\Snr}}{Supernova remnant}
\newcommand{{\snr}}{supernova remnant}
\newcommand{{\snrs}}{supernova remnants}
\newcommand{{\Snrs}}{Supernova remnants}
\newcommand{{\sn}}{supernova}
\newcommand{{\sne}}{supernovae}
\newcommand{{\pwn}}{pulsar wind nebula}
\newcommand{{\pwne}}{pulsar wind nebulae}
\newcommand{{\dsa}}{DSA}
\newcommand{\arcmin}{{$^{\prime}$}}

\DeclareMathAlphabet\mathbfcal{OMS}{cmsy}{b}{n}

\def\idxg11{\index{G11.2-0.3}}

\def\idx79j{\index{SN 1979C}}

\renewcommand{\S}{Sect.}

\begin{document}


\title*{Nonthermal processes and particle acceleration in supernova remnants}
\author{Jacco Vink\thanks{corresponding author} and Aya Bamba}
\institute{Jacco Vink \at Anton Pannekoek Institute for Astronomy \& GRAPPA, University of Amsterdam, Science Park 904, 1098 XH Amsterdam, The Netherlands \&\\  
SRON National Institute for Space Research, Niels Bohrweg 4
2333 CA,  Leiden, The Netherlands
\email{j.vink@uva.nl}
\and
Aya Bamba \at The University of Tokyo, Hongo 7-3-1, Bunkyo-ku, Tokyo, 113-0033, Japan \&\\
Research Center for the Early Universe, School of Science, The University of Tokyo, 7-3-1
Hongo, Bunkyo-ku, Tokyo 113-0033, Japan
\email{bamba@phys.s.u-tokyo.ac.jp}}

%
%


\maketitle
\abstract{%
Shocks of supernova remnants (SNRs) accelerate charged particles
up to $\sim$100~TeV range
via diffusive shock acceleration (DSA) mechanism.
\index{diffusive shock acceleration theory}
It is believed that shocks of SNRs are the main contributors to the pool of  Galactic cosmic rays,
although it is still under debate whether they can accelerate particles
up to the ``knee"  energy ($10^{15.5}$~eV) or not.
In this chapter, we start with introducing SNRs as likely sources of cosmic rays and the radiation mechanisms associated with cosmic rays
(section 3). 
In the section 4, we summarize the mechanism for particle acceleration,
including basic diffusive shock acceleration and  nonlinear effects, as well as discussing the injection problem.
Section 5 is devoted to the X-ray and gamma-ray observations of
nonthermal emission from SNRs, and what these reveal about the
cosmic-ray acceleration properties of SNRs.
}


\section{Keywords} 
cosmic rays, supernova remnants, 
shocks, diffusive shock acceleration (DSA),
magnetic fields, synchrotron, inverse Compton scattering, 
pion decay

\section{Introduction}
In a series of seminal papers, Baade \& Zwicky hypothesised
that the most luminous novae---which they dubbed ``super novae''---marked
the transition of a massive star into a neutron star.
They also made the suggestion that supernovae (SNe) are responsible for
accelerating the highly energetic particles that reach us from outside
the solar system, the so-called cosmic rays \citep{baade34}.

Since the 1930s our understanding of both SNe and cosmic rays have greatly expanded.
We know now that there are both neutron star producing core-collapse SNe as well as thermonuclear
SNe (Type Ia SNe). It has  been established  that cosmic rays consist for 99\%  of atomic nuclei, with
sources in the Milky Way responsible for proton cosmic rays up to $\approx 3\times 10^{15}$~eV---below these energies they are referred to as Galactic cosmic rays. And likely only
above $10^{17}$--$10^{18}$~eV are cosmic rays originating from outside the Galaxy.
But the idea that SNe are somehow responsible for most of the Galactic cosmic rays is still  a leading hypothesis.

The SN hypothesis for the origin of cosmic rays is supported by the match between SN energetics and the energy budget required for cosmic ray production,
which require that about 5--10\% of SN explosion energy is used
for acceleration cosmic rays.
However, the main evidence for cosmic-ray acceleration powered by SNe has been the detection of nonthermal emission from SN remnants (SNRs).
In this chapter we will review the X-ray and gamma-ray evidence for this hypothesis. Initially the evidence
 consisted of the detection of synchrotron  radio emission from SNRs, caused by the presence of  relativistic electrons inside SNRs \citep{shklovsky54}.
Although electrons constitute $\sim 1$\% of the cosmic rays locally detected, the synchrotron radiation  provided evidence that charged
particles have somehow been accelerated to relativistic energies. The identification of synchrotron radiation from SNR shells also suggested that cosmic-ray  acceleration may be powered
by SN explosions, but that the acceleration itself is occurring around the shocks caused by the explosion in the first few hundreds to thousands of years after
the SN explosion. A theory for how shocks accelerate charged particles was developed in the late 1970s \citep{axford77,krimsky77,bell78a,blandford78}, and is known as
diffusive shock acceleration  (DSA) or first order Fermi acceleration \citep{malkov01}.
\index{Fermi acceleration process}

The idea that  SNRs, rather than SNe, are accelerating cosmic rays  received a major boost due to the discovery of nonthermal X-ray emission from SNRs in 1995
\citep{koyama95}---later followed by the detection of TeV gamma-ray emission from  SNRs \citep{aharonian01,aharonian04}
after some earlier hints at lower gamma-ray energies \citep{esposito96}.

The nonthermal X-ray emission is caused by synchrotron emission from electrons with
$\gtrsim 10$~TeV in energy, and it, therefore, testifies of the ability of shocks to accelerate charged particles to ultra-relativistic energies. Moreover, since these electrons
radiatively lose their energies with the lifetime of young SNRs, it identifies the SNR stage, rather than the SN event itself, as the period in which cosmic rays are accelerated.

The detection of gamma-ray emission in the very-high energy  (VHE) gamma-ray regime ($\gtrsim 100$~GeV) confirms the presence of ultrarelativistic particles in SNRs.
In addition, the broad-band gamma-ray emission from $\sim$200~GeV to 100~TeV  provides evidence that at least in some cases the gamma-ray emission is caused by
relativistic ions. So gamma-ray emission currently provides the best evidence for the acceleration of atomic nuclei in SNRs---usually referred to as hadronic cosmic rays---which
make up $\sim$99\% of the cosmic rays detected locally.

This is not to say that the case for a SNR origin for Galactic cosmic rays has been settled now. The current evidence does not support the idea that SNRs are capable
of acceleration cosmic rays up to the cosmic-ray ``knee"  at $\sim 3\times 10^{15}$~eV. Solving this puzzle will be a challenge for the next two decades.
Three leading ideas currently exist: 1) perhaps the interpretation of the cosmic-ray ``knee"  as being the maximum energy for proton cosmic-ray acceleration needs to be revised; 2)
the highest energy cosmic rays are accelerated in the first year up to 10--100~yr after explosion of all supernovae; or 3) plasma turbulence in starforming regions, 
powered by SNe and stellar winds collectively are responsible for the highest energy cosmic rays.

In this chapter we will focus on nonthermal X-ray and gamma-ray emission from SNRs and the implications for cosmic-ray acceleration by SNR shocks, but will
also discuss the need for some alternative ideas, which includes recent evidence for PeV cosmic rays in starforming regions.\footnote{
We note that some SNRs contain an energetic pulsar, which creates
a pulsar wind nebula that also emits nonthermal radiation. Here we concentrate, however, solely on the nonthermal emission from SNRs shells.}

\section{SNRs as the origin of Galactic cosmic rays}
Around 1911 Victor Hess \citep{hess12}  established, using balloon flights, that ionising radiation comes from outside the Earth atmosphere. Soon after this ``penetrating radiation" was referred to as ``cosmic rays", a bit of a misnomer as in the 1930s it became
clear that cosmic rays consist of energetic ($\gtrsim 0.1$~GeV) charged particles that reach Earth from outside the solar system.

\subsection{The cosmic-ray spectrum}
The cosmic-ray energy spectrum is now well measured from $\sim 10^8$~eV up to $\sim 10^{20}$~eV, and the distribution shows a power-law
spectrum  $\propto E^{-q}$, with $q\approx 2.7$. However, the spectrum steepens around $3\times 10^{15}$~eV---a feature known as the ``knee" ---and flattens around $3\times 10^{18}$~eV (the ``ankle"), around $5\times 10^{19}$~eV the steepens again.
For this chapter we are mainly concerned with the connection of cosmic rays with energies below the ``knee" ,  as it is thought that
the``knee" corresponds to the maximum energy protons are accelerated to in Galactic cosmic-ray sources. There is evidence
from air-shower measurements that the cosmic-ray composition becomes heavier above the ``knee"  \citep{kascade11}. The "ankle" then likely
marks the division between Galactic cosmic rays, and cosmic rays of extragalactic origin.

{\em
So if indeed SNRs are the main sources of Galactic cosmic rays, they must be  able to accelerate protons up to the ``knee"  at 
$3\times 10^{15}$~eV---3 PeV---which is challenging from a theoretical point of view \citep{lagage83}. Sources of $>$PeV 
cosmic rays are often referred to as {\em PeVatrons}. So one of the leading unanswered questions  is whether SNRs are indeed PeVatrons,
and if not, what are the Galactic PeVatrons responsible for cosmic rays up to the ``knee" ?}

\subsection{The cosmic-ray composition and leptonic versus hadronic cosmic rays}
Most cosmic rays ($\sim 99$\%) are protons or heavier atomic nuclei, collectively labeled as hadronic cosmic rays. 
\index{hadronic cosmic rays}
\index{cosmic rays}
Below $\sim 100$~GeV the composition of cosmic rays can be directly measured using detectors carried aboard balloon flight or satellites. This shows that the abundance pattern of cosmic-ray particles is similar to the solar-system abundance pattern, except
that the odd-Z elements, such as boron ($Z=5$), are more abundant than in the solar system, as a result of the break-up of cosmic-ray nuclei due to collisions with interstellar gas. The ratio between odd and even Z elements provides information on the typical 
path lengths
of cosmic ray particles from their sources of origin till detection.  Moreover, the presence of radioactive nuclei, such as
$^{10}$Be, $^{26}$Al, $^{36}$Cl, and $^{54}$Mn can be used to estimate the amount of time that cosmic-ray particles spend on average
in the Milky Way. From this it is inferred that cosmic rays with energies around 1~GeV spend about $\tau_{\rm esc}\approx 1.5\times 10^7$~yr in the Milky Way, and that they must reside in low density regions for a large fraction of this time. 
Given that the cosmic rays are relativistic, the long escape time suggests that the particles move diffusively through the interstellar medium (ISM), characterised by a diffusion coefficient $D\approx 4 \times 10^{28}$~cm$^2$\,s$^{-1}$.
The diffusion coefficient is energy dependent, with a proportionality of $\propto E^\delta$, with $\delta\sim 0.3$--$0.7$ \citep{strong07}.

The abundance pattern, unfortunately, contains little information about the acceleration sites. However, there is evidence for an overabundance of $^{22}$Ne \citep{binns05}, indicative of stellar-wind enriched material, and for a slight overabundance of volatile
elements, perhaps indicative of dust particles injected preferentially into the acceleration process \citep{meyer97}.

Only $\sim 1$\% of the Galactic cosmic rays are electrons, and even a smaller part are positrons---together often labeled ``leptonic cosmic rays". However, electrons, having a lower inertial mass than hadrons, are radiatively much more efficient than hadrons. \index{leptonic cosmic rays}
As we will describe in Sect.~\ref{sec:radiation}, relativistic electrons produce 
radiation from the radio band to X-rays in the form of synchrotron radiation, 
but also produce gamma rays, mostly through inverse Compton scattering.
Hadronic cosmic rays only significantly contribute to the gamma-ray part of the electromagnetic spectrum, making  gamma-ray studies essential for
studying the cosmic-ray acceleration efficiency
of SNRs.

\subsection{The Galactic cosmic-ray energy budget}
Although associating SNRs with Galactic PeVatrons is challenging from both a theoretical \citep{lagage83} and observational
point of view (as we will see), the main reason why SNRs and their associated energy source---SN explosions--are compellingly
associated with Galactic cosmic rays is the available energy budget.
The local energy density of cosmic rays is $U_{\rm cr}\sim 1$~eV\,cm$^{-3}$. If we approximate the volume of the Galaxy that is occupied by cosmic rays 
with a thick disc with radius of $\sim$10~kpc and thickness of 2~kpc, we have a total
energy in cosmic rays of $E_{\rm cr}\approx U_{\rm cr}V\approx 3\times 10^{55}$~erg. Since cosmic rays have a typical
escape time from the Galaxy of 15~Myr, the power needed for maintaining the cosmic-ray energy density  in the Galaxy is 
$\dot{E}_{\rm cr}\approx E_{\rm cr}/\tau_{\rm esc}\approx 6\times 10^{40}$~erg\,s$^{-1}$.
This compares well with the total power provided by SN explosions. For 2 explosions per century, and $10^{51}$~erg of explosion energy,
we find $\dot{E}_{\rm SN}\approx 6\times 10^{41}$~erg\,s$^{-1}$. So an cosmic-ray efficiency is needed for $\approx 10$\%.
The total stellar wind power in the Galaxy is less, and also the rotational energy provided by pulsars is much less if they
are born with initial spin periods $P_0>5$~ms, as seems to be generally the case \citep[e.g.][Chapter 12]{vinkbook}.

\subsection{Radiation from leptonic and hadronic cosmic rays}
\label{sec:radiation}

So far we dealt with information about cosmic rays obtained by directly, or indirectly measure the properties of cosmic rays that reach the solar system. However, for measuring the properties of cosmic rays in cosmic accelerators, here SNRs, we rely on the radiation from cosmic rays while still being in their sources. Since the cosmic-ray spectrum is a nonthermal distribution, the ensuing radiation is  often referred to as ``nonthermal radiation". But it should be
noted that the radiation mechanisms responsible occur irrespective of whether the distribution is thermal
or nonthermal, it is just that certain radiation types are predominantly associated with nonthermal distributions.

\subsubsection{Synchrotron radiation}
\index{synchrotron radiation}
Historically the first type of emission that was associated with nonthermal particle 
distributions is synchrotron radiation, which was first 
identified in astrophysical context in the radio bands \citep{shklovsky54}. 
Synchrotron radiation occurs when charged particles
are forced to helical path following magnetic field lines, due to the Lorentz force. The changing electric-field
associated with the non-linear part of the path results in a radiation pattern that is mostly directed in the
direction perpendicular to the magnetic-field direction. The power emitted by a single particle scales with its
mass as $m^{-2}$, which makes that synchrotron radiation from electrons is $3\times 10^6$ stronger than from protons.
So synchrotron radiation is generally associated with either electrons or positrons, i.e. leptonic cosmic rays.
The typical frequency at which an electron emits synchrotron radiation is  
\begin{align}
\nu_{\rm syn} \approx & 0.29 \times \frac{3}{2} \gamma^2 \frac{eB_\perp}{m_{\rm e} c} = 1.8\times 10^{18}E^2B_\perp~{\rm Hz} \label{eq.sync-spec}\\\nonumber
\approx & 0.47\times 10^9\left(\frac{E}{1~{\rm GeV}}\right)^2\left(\frac{B_\perp}{100~{\rm \mu G}}\right) ~{\rm Hz}
\end{align}
with $\gamma$, $E$ respectively the Lorentz factor and energy (in erg, unless otherwise indicated) of the electron, and $B_\perp$ the perpendicular
component of the magnetic-field strength.
The factor 0.29 introduced corresponds to the peak of the broad spectral radiation distribution from
a mono-energetic electron.  This shows that radio synchrotron radiation around 1~GHz requires the presence
of electrons with energies well in excess of a GeV, for magnetic fields that are typical for the interstellar
medium of $\sim 5~{\rm \mu G}$.

The radiated power from a single electron can be expressed as
\begin{equation}\label{eq:syn_losses}
    \left(\frac{dE}{dt}\right) = \frac{4}{3}\sigma_{\rm T}c\beta^2\gamma^2 U_{\rm B},
\end{equation}
with $\sigma_{\rm T}=6.65\times 10^{-25}$~cm$^2$ the Thomson cross section, $\beta=v/c$ (electron speed in units of the speed of light), and $U_{\rm B}=B^2/8\pi$ the magnetic-field energy
density. This also sets the timescale for an electron to lose its energy due to synchrotron radiation:
\begin{equation}
    \tau_{\rm syn} \approx \left|\frac{E}{dE/dt}\right|
    = \frac{9}{4}\frac{(m_{\rm e}c^2)}{e^4cB^2E}
    \approx \frac{634}{B^2E}~{\rm s}\cong 1258~\left(\frac{E}{1~{\rm TeV}}\right)^{-1}
    \left(\frac{B}{100~{\rm \mu G}}\right)^{-2}~{\rm yr}.
\end{equation}

The timescale suggests that for radio synchrotron emitting electrons around a few GeV, radiative losses are not
important during typical lifetimes of SNRs ($\sim 10^5$~yr). However, for X-ray synchrotron emitting electrons,
which are typically $\gtrsim 10$~TeV, radiative losses are important. See Sect.~\ref{sec:xray}.
For SNRs synchrotron radiation from young SNRs has been detected up to $\sim 150$~keV \citep[e.g.][]{the96,green15}.

For nonthermal distributions of electrons characterised by an
energy distribution of the form $N(E)dE\propto E^{-q}dE$, the resulting synchrotron emissivity
spectra have a power-law distribution of $\epsilon_{\rm syn}(\nu)d\nu \propto \nu^{-\alpha} d\nu$, with $\alpha = (q-1)/2$. (Note that in the radio astronomy
literature  the minus sign is often included in the definition of $\alpha$. )
This relation follows from the factor $\gamma^2$ in (\ref{eq:syn_losses}).
In
high-energy astrophysics it is more common to use the photon number emissivity
($n_\nu = \epsilon_\nu/h\nu$), resulting in a power-law index for the photon spectrum
of $\Gamma=\alpha+1=(q+1)/2$.

Synchrotron radiation is intrinsically polarized, with
the electric vector of the polarization being perpendicular to the magnetic-field direction.
The maximum possible polarization fraction from  nonthermal electron distribution is
\begin{equation}
    \Pi = \frac{\alpha+1}{\alpha+5/3},
\end{equation}
corresponding to 69\% for $\alpha=0.5$. In reality the
polarization fraction is usually much lower, due to different magnetic-field orientations along the line of sight, or due magnetic-field turbulence. Nevertheless,
radio-synchrotron polarization is often used to identify radio emission as synchrotron radiation.
\index{synchrotron polarization}

\subsubsection{Inverse Compton scattering}
\label{sec:ic}
\index{inverse Compton scattering}

The same relativistic electrons that  produce synchrotron radiation can also scatter background photons through
inverse Compton scattering. Inverse Compton scattering can be thought of as Thomson scattering of a photon
in the rest frame of the electron. In this frame a background photon with energy $h\nu_0$ has a Lorentz
boosted energy of $h\nu^\prime\approx \gamma(h\nu_0)$ (with some angle dependence). 
Thomson scattering does not change the energy of the photon.  
However, transforming back to the
observer frame results in an additional Lorentz boost, but with an energy distribution that depends
on the scattering angle with respect to the motion of the electron. So the final scattered photon appears to
have an energy of the order of $h\nu^{\prime\prime}\approx \gamma^2 h\nu_0$ \citep[see e.g.][for details]{vinkbook}.
As an example, an electron with energy of 1~TeV will upscatter a typical cosmic-microwave background (CMB) photon of 0.0007~eV to an energy of 2.7~GeV.

For a single electron the radiation power is
\begin{equation}
 \left(\frac{dE}{dt}\right) = \frac{4}{3}\sigma_{\rm T} \gamma^2 c\int h\nu_0 n_{\rm rad}(\nu_0)d\nu_0 = \frac{4}{3}\sigma_{\rm T} \gamma^2 c U_{\rm rad},
\end{equation}
with $U_{\rm rad}$ the radiation energy density. One recognises here that $\sigma_{\rm T} c n_{\rm rad}(\nu_0)$ is 
the collision rate between electrons and background photons with a  frequency $\nu_0$. 
Note that the equation, as written on the
right hand side, is similar to that of synchrotron radiation, except that the magnetic-field energy density
factor $U_{\rm B}$ is replaced by the radiation energy density---c.f. Eq.~\ref{eq:syn_losses}.

As the peaks of an spectral energy distribution (SED) indicates approximately 
how much flux or luminosity is produced as a function of frequency. Therefore, comparing the peaks in the SED of
the synchrotron component (radio to X-rays) with the SED of  gamma-ray inverse Compton scattering  
provides  an estimate
of the ratio of the energy density in background radiation and magnetic field:
$P_{\rm syn}/P_{\rm IC}\approx U_{\rm B}/U_{\rm rad}$.

An omnipresent component of the radiation energy density in the Universe is 
that associated with the cosmic microwave background (CMB), which is $U_{\rm rad,CMB}\approx 0.26$~eV~cm$^{-3}$. In addition, Galactic infrared emission and UV stellar light 
can have additional contributions that may be as large as the CMB, depending
on the local environment of a source.

The cross section for inverse Compton scattering is seriously reduced if the
photon energy in the rest frame of the electron approaches or exceeds 
the rest mass energy,
i.e. $h\nu^\prime = \gamma h\nu_0 \gtrsim m_{\rm e} c^2$. For the scattered photon
energy in the frame of the observer this means that for 
$h\nu^{\prime\prime}\gtrsim (m_{\rm e}c^2)^2/h\nu_0$ the inverse Compton radiation component is strongly suppressed. This corresponds to $h\nu^{\prime\prime}\gtrsim 400$~TeV for the typical CMB photon energy of 0.0007~eV. However, for a typical
optical or UV of around 1~eV suppression starts already around upscattered photon
energies of 0.2 TeV. This suppression of inverse Compton scattering and related effects are usually referred to as Klein-Nishina suppression/effects.

Apart from Klein-Nishina effects, the spectral index of the photon spectrum is similar
to that of synchrotron radiation: $\Gamma=(q+1)/2$.

\subsubsection{Nonthermal bremsstrahlung}
\label{sec:bremss}
\index{bremsstrahlung}

Bremsstrahlung (or free-free emission) is the result of the (small) deflection of a charged particle when it passes close to another
charged particle. Since the deflection depends on the mass of the particle, and electrons are light, usually the dominant
form of bremsstrahlung is electron bremsstrahlung.
Often  bremsstrahlung from hot astrophysical plasmas are caused by electrons with a thermal (Maxwellian) velocity distribution. Hence, the name thermal bremsstrahlung. 
The total emissivity of thermal bremsstrahlung scales as
$\epsilon_{\rm ff} \propto n_{\rm e} n_{\rm p} T_{\rm e}^{1/2}$. The differential emissivity (i.e. per unit energy or frequency) scales  as $\epsilon_{\rm ff,\nu} \propto n_{\rm e} n_{\rm p} T_{\rm e}^{-1/2}\exp\left(-h\nu/kT_{\rm e }\right)$.

However, particle acceleration also results in a population of nonthermally distributed electrons. At high energies these are referred to as leptonic cosmic rays. At lower energies, however, nonthermal bremsstrahlung may arise from the suprathermal electron population that constitutes
the injection spectrum of the electron cosmic-rays, or more in general, a nonthermal leftover from collisionless shock physics. Note that collisionless shocks may not immediately lead
to a thermal distribution of particles. 

The suprathermal electrons should give rise to  nonthermal, hard X-ray continuum radiation, best identified above 10~keV. The reason is that the X-ray below 10~keV SNRs is often dominated
by thermal radiation processes, be it line emission, or thermal bremstrahlung continuum and related processes, such as free-bound emission.
Indeed, \cite{asvarov90} argued that the hints of hard X-ray continuum from young SNRs like Cas A and Tycho, are formed by the low-energy tail of the electron cosmic-ray population.
When hard X-ray emission from Cas A and other young SNRs was firmly established \citep{the96,allen97,favata97} there was a debate about the nature of this emission:
was it nonthermal bremsstrahlung \citep[e.g.][]{laming01a} or synchrotron radiation \citep{allen97,reynolds98}? 
The latter caused by electrons from near the maximum of the electron cosmic-ray spectrum ($\sim$10--100~TeV).
Although the debate has not been completely settled, the general concensus is that for young SNRs the hard X-ray emission forms the high-energy tail of the synchrotron component. See
for example the discussion on the NuSTAR imaging spectroscopy of the hard X-ray emission from Cas A \citep{grefenstette15}. 

One reason why a nonthermal
bremsstrahlung interpretation has difficulties is that Coulomb collisions between the thermal plasma and the suprathermal electrons results in significant energy exhanges at low electron energies, and it leads
to energy losses for suprathermal electrons. The energy-loss rates scale for nonrelativistic electrons as $dE_{\rm e}/dt \propto n_{\rm H} E^{-1/2}$. This means that a nonthermal electron distribution at the shock front
quickly starts losing its lower electron energy part, resulting in an inverted electron spectrum
\citep{vink08a}. The timescale on which this happens for electrons below 100 keV is  $\tau \approx 10^{11}~n_{\rm e}^{-1}$~s,
which is the timescale applicable to Cas A. However, for SN1006, with its lower density plasma, nonthermal bremsstrahlung may be relevant.

Although nonthermal X-ray bremsstrahlung is not discussed so often anymore for young SNRs, 
recently evidence based on NuSTAR observations shows that the hard X-ray emission from SNR W49B  is relatively hard, with a power-law
number index of $\Gamma\approx 1.4$. It likely constitutes a case of nonthermal bremsstrahlung \citep{tanaka18}. This can be verified in the future,
as a nonthermal electron population with energies in the 10--100 keV range leaves an imprint on   Fe-K line ratios, which can be measured by the next generation calorimetric X-ray detectors on board XRISM
and Athena.
\index{NuSTAR}

Nonthermal bremsstrahlung as an ingredient for the (very) high-energy gamma-ray emission is less controversial than for the X-ray band. The responsible electrons (and possibly positrons)
are part of the cosmic-ray population, and will surely contribute to the gamma-ray emission. As the emission is caused by electrons, bremsstrahlung is labeled a leptonic emission component,
together with inverse Compton scattering (Section~\ref{sec:ic}). Like the thermal bremsstrahlung component,  nonthermal bremsstrahlung emissivity scales with density as $\epsilon_{\rm ntb}n_{\rm e,cr}n_{\rm p}$, with  $n_{\rm e,cr}$ referring to the
relevant leptonic cosmic-ray density. This scaling is identical to pion decay emission, see Section~\ref{sec:pions} below. Moreover, inverse Compton radiation is the dominant
leptonic gamma-ray component for densities $n_{\rm p} \lesssim 240$~cm$^{-3}$ \citep{hinton09}, even when only considering the cosmic-microwave background as a source of seed photons.
For  modeling the SEDs of SNRs  it is usually assumed that the
electron over proton cosmic-ray number ratio is  $n_{\rm e}/n_{\rm p} \sim 1/100$. In that case pion decay dominates the gamma-ray emission above $\sim 100$~MeV over bremsstrahlung.
For these reasons nonbremsstrahlung is often neglected when modeling broad spectral energy distribution in gamma-rays: for low densities inverse Compton scattering dominates over bremsstrahlung, whereas
for high densities pion decay dominates over bremsstrahlung.

\subsubsection{Pion production and decay}
\label{sec:pions}
\index{pions}
\index{pi-mesons}
\index{pion decay}

The  radiation mechanisms discussed above are mostly associated with  leptonic
cosmic rays (electrons/positrons). Only one mechanism is associated with
hadronic cosmic rays: pion decay. Pions are the lightest mesons (hadrons with two quarks)
and come in three flavors: $\pi^0$,$\pi^+$, and $\pi^-$. They are produced 
whenever a hadronic cosmic-ray particle (usually a proton) collides with
target hadron, usually a proton in the local  gas.
For the lightest pion, $\pi^0$ ($m_{\pi^0}c^2=135$~MeV), a threshold energy
of 280~MeV for the cosmic-ray proton is needed for the reaction 
$p+p\rightarrow p+p + \pi^0$. For a charged pion a reaction like
$p+p\rightarrow p+n + \pi^+$ is needed. At sufficiently high energies multiple
pions, and even new protons and neutrons can be created, provided nature's conservation
laws, such as energy, momentum, charge, and baryon number are obeyed.

Pions are unstable particles and decay into lighter particles and photons:
\begin{align*}
\pi^0  \rightarrow & 2\gamma, \\
\pi^+  \rightarrow & \mu^+ + \nu_\mu\\
                   & \downarrow\\ \nonumber
                   & e^+ + \nu_{\rm e} + \bar{\nu}_\mu, \\\nonumber
\pi^-  \rightarrow & \mu^- + \bar{\nu}_\mu\\
                   & \downarrow\\ \nonumber
                   & e^- + \bar{\nu}_{\rm e} + \nu_\mu.  \nonumber
\end{align*}
So one of the outcomes of charged pion decay are electron and muon
(anti)neutrinos, which are the particles of interest for high-energy 
neutrino detectors like IceCube and the future KM3NeT. No high-energy neutrino signals have yet
been detected from SNRs.
\index{neutrinos}
\index{IceCube}
\index{KM3NeT}

\begin{figure}
\centerline{
\includegraphics[width=0.9\textwidth]{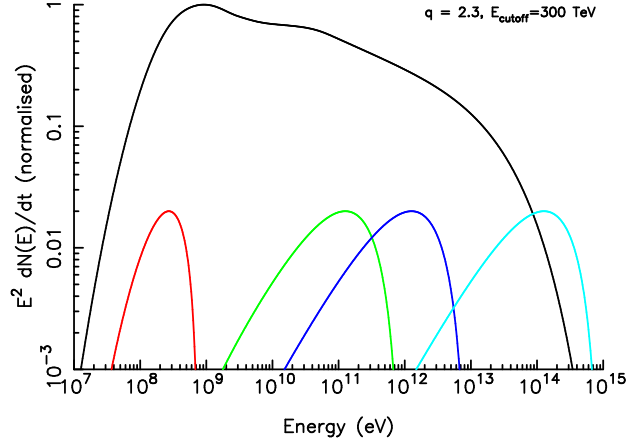}
}
\caption{\label{fig:pions}
A model SED for a pion decay gamma-ray spectrum based on the semi-analytical functions of \citep{kafexhiu14}. The input proton spectrum is assumed to have a power-law distribution in momentum $n(p)p^2pdp \propto p^2p^{-q-2}$, with $q=2.3$. 
The colored lines show the results for monochromatic ``beams" of protons with
energies of   1 GeV (red), 1 TeV (green), 10 TeV (blue), and 1 PeV (cyan).
The low energy cutoff, together with the peak around $5\times 10^8$~eV is often
referred to as the ``pion bump".
(Based on a figure in \citep{vinkbook}.)
}
\end{figure}

For this chapter it is the neutral pion production that is of interest, as
neutral pion decay leads to two photons each with an energy of $\frac{1}{2}m_{\pi^0}c^2=67.5$~MeV in the restframe of the pion. However, a high
energy cosmic ray ($E\gg 280$~meV) will lead to a pion with a considerable
kinetic energy. So the  gamma-ray photons from the decay will receive a Lorentz
transformation, resulting in a photon energy
of $h\nu= \frac{1}{2}m_{\pi^0}\gamma_{\pi^0}(1 \pm \beta_{\pi^0}\cos \theta)$,
with $\theta$ the angle between the pion  direction and the direction of the
two decay photons, which move in opposite directions in the frame of the pion.
So each decay results in a low energy photon---associated with the minus sign---
and a high energy photon---associated with the plus sign.
Details of the relation between cosmic-ray energy and the expected probability distribution of the photon energies can be found in, for example, \cite{kafexhiu14}.
On average the higher energy photon has an energy that is about 10\% of the energy
of the cosmic-ray proton.
See Fig.~\ref{fig:pions} for a typical SED for emission caused by hadronic
cosmic rays, including the distribution in energy for protons with a give energy.
For a proton cosmic-ray spectrum of $N(E)dE \propto E^{-q}$ the resulting
photon spectrum has a similar spectral index, i.e. $\Gamma\approx q$.
We expect, therefore, that a gamma-ray spectrum dominated by pion decay has a spectral
index of $\Gamma \approx 2.2$ (for $q=2.2$), whereas for the inverse Compton spectrum
from a population of electrons with the same spectral index $q$ the photon spectrum is harder: $\Gamma \approx 1.6$.

A defining characteristic of a pion decay spectrum is that the spectrum is symmetric
in logarithmic energies around half the rest mass energy of the pion, due to the $\pm$
sign in the Lorentz transformation of the two oppositely moving photons. In an SED,
for which the spectrum is multiplied by $E^2$, 
this results in a characteristic peak around 100-200~MeV and a sharp cutoff below
these energies. This is referred to as the ``pion bump" (see also Fig.~\ref{fig:pions}).
\index{pion bump}

\section{The mechanism of diffusive shock acceleration}

\subsection{Collisionless shocks}
\index{collisionless shocks}
Particle acceleration appears to be a common byproduct of so-called collisionless shocks. In collisionless shocks the changes in thermodynamic properties of the plasma across the shock boundary are not caused by particle-particle interactions through Coulomb interactions, but through collective effects---particles are randomly deflected by magnetic-field turbulence and/or electric fields.
Collissionless shocks are common in many astrophysical situations as the densities are low, making Coulomb interactions rare.
The absence of efficient Coulomb interactions does not allow
charged particles that have velocities greatly in access of the local
plasma velocity to redistribute their excess energy to other plasma particles, allowing them to  acquire even more energy.
In a collisionless shock it is also not necessary for the electrons and ions to equilibrate their temperatures. Apart from these consideration, 
the standard Rankine-Hugoniot shock equations can be used, which state that mass-, momentum-, and enthalpy-flux across the shock
boundary are conserved. For a strong shock--- i.e. $M=V_{\rm s}/c_{\rm s,1} = \sqrt{P_1/\rho_1 V_{\rm s}^2}\gtrsim 5$ with $c_{\rm s}$ the sound speed in the unshocked plasma---the post-shock plasma is characterised by
\index{Rankine-Hugoniot equations}
\begin{align}
\chi \equiv & \frac{\rho_2}{\rho_1} = \frac{(\gamma+1)}{(\gamma-1)}=4~~(\mbox{\rm for}~\gamma=5/3),\\
<kT_2> =&\frac{3}{16}\mu V_{\rm s}^2,\\
v_2= &\frac{1}{\chi}V_{\rm s}=\frac{1}{4}V_{\rm s},\\
\Delta v=& v_1 - v_2 = \left(1-\frac{1}{\chi}\right)V_{\rm s}=\frac{3}{4}V_{\rm s},
\end{align}
with subscript 1 and 2 indicating upstream (unshocked) and downstream (shocked) quantities,
$V_{\rm s}$ the magnitude of the shock velocity, $v_1$ and $v_2$ the plasma velocities in the frame of the shock (i.e. $|v_1|=V_{\rm s}$), $\gamma$ the plasma's adiabatic index, $\chi\equiv \rho_2/\rho_1$ the shock compression ratio, and $\mu$ the average particle mass---$\mu\approx 0.6$ for solar system abundances.

The strong shock approximation ($M\gtrsim 5$) is well justified for young SNRs,
which have $V_s\approx 1000$--$7000~{\rm km\,s^{-1}}$, 
even if they are evolving in the hot part of the interstellar medium, where the typical sound speed is $c_{\rm s}\sim 10~{\rm km\,s^{-1}}$. However, once 
the shock velocity has dropped below $\approx 30~{\rm km\,s^{-1}}$, the strong shock
approximation breaks down, also because the typical turbulent gas motions are characterised by similar velocities. Shock velocities below 50~km\,s$^{-1}$ 
occur typically for SNRs that are several ten thousand years old.
Note that for $V_{\rm s}\lesssim  200~{\rm km\,s}^{-1}$---typically happening for SNR ages of 5000--20000~yr--- post-shock radiative cooling becomes
important, due to strong optical/UV radiation. This results in very high shock compression ratios further downstream of the shock, 
as the plasma tries to maintain in pressure
equilibrium, in the presence of energy losses. The high shock compression ratios
leads to the optical narrow filaments that characterises the optical emission from
middle-aged SNRs like the Cygnus Loop (Veil nebula) or Spaghetti Nebula (Simeis 147).
For more details on this topics, see  this handbook in the Chapter on ``Supernova remnants: types and evolution” by  by A. Bamba \& B. Williams.

\subsection{Diffusive shock acceleration theory and its extensions} 
Efficient acceleration likely  mostly in young SNRs through the diffusive shock
acceleration (DSA) mechanism \citep{bell78a,axford77,krimsky77}.
DSA operates on charged particles with energies well above $kT_2$. These particles
have speeds greatly in access of $V_{\rm s}$ and are elastically scattering due to magnetic-field irregularities. As a result they
diffusively wander through the plasma, which brings them every now and then across the shock front. The magnetic-field irregularities,
probably associated with Alfv\'en or magneto-acoustic waves are on average at rest with the local plasma, which moves with $v_1$ in the
unshocked plasma, and with $v_2$ in the shock medium.  The
assumption is that for non-relativistic shocks the scattering
of energetic charge particles makes the velocity distribution
isotropic after a few scatterings.
Every time a particle crosses the shock front it has an excess velocity
of $\Delta v$ with respect to the new frame of reference, and the particle appears to have an excess energy given
by a Lorentz boost ($E\rightarrow E^\prime$). If we take the particle speed to be relativistic ($v\approx c$), the Lorentz boost of $E^\prime=\Gamma(E-\Delta v/c \cos\theta)$, with a non-relativistic boost $\Gamma\approx 1$, the gain in energy is
\begin{equation}
   \frac{ <\Delta E>}{E}\approx \left< \left|\frac{\Delta v}{c}\cos \theta \right|\right>\approx \frac{2}{3}
\frac{V_{\rm s}}{c}\left(1-\frac{1}{\chi}\right).
\end{equation}
A full cycle, in which the particle crosses from downstream to upstream and back, provides twice this
energy. Since the energy gain is proportional to the energy itself, the energy of a charged particle
growth exponentially with the number $K$ of shock crossing cycles:
\begin{equation}\label{eq:ek}
E_K= E_0\left[ 1 +  \frac{4}{3}
\frac{V_{\rm s}}{c}\left(1-\frac{1}{\chi}\right)\right]^K,
\end{equation}
with $E_0$ some initial energy.

On the other hand, once particles are too far removed from the shock
front downstream of the shock, 
they are unlikely to recross the shock front again. Once this
is the case the final energy is frozen in, and the particle
is said to have escaped downstream---upstream escape is unlikely
as the plasma will eventually be swept up by the shock. However,
at very high energies, upstream escape can and probably does occur.
The chance of escaping downstream during one shock cycle
depends on the ratio between the plasma velocity $v_2$ and the
particle velocity (assumed to be $c$):
\begin{equation}
    P_{\rm esc} =4 \frac{v_2}{c}= 4 \frac{1}{\chi}\frac{V_{\rm s}}{c},
\end{equation}
with the factor $4$ arising from averaging over the angle
of particle velocity with respect to the shock. So if
inject $N_0$ particles with energy $E_0$ into the DSA process,
after $K$ full cycles the number of particles has been
reduced to
\begin{equation}
N(E>E_K) = N_0 (1-P_{\rm esc})^K,
\end{equation}
with $E_K$ given by (\ref{eq:ek})
The combination of exponential growth in energy with an exponential
decay in the number of particles that are still
participating in  the DSA process leads to a power-law distribution:
\begin{align}
    N(>E) = C E^{-q+1} \rightarrow \frac{dN(E)}{dE} = C^\prime E^{-q},
\end{align}
with 
\begin{equation}\label{eq:dsa_index}
    q = \frac{\chi+2}{\chi-1} = 2 ~(\mbox{for}~ \chi=4).
\end{equation}
Note that another way to derive this result is based on phase-space (momentum-space) conservation in the absence of collisions \citep{blandford78,drury83}, which gives
a momentum distribution of
\begin{equation}
    N(p)4\pi p^2 dp \propto  p^{-q+2}dp,
\end{equation}
which is identical to $N(E)\propto E^{-q}$, but does not require the relativistic approximation $E=|p|c$ and indicates that the expected cosmic-ray is
characterised by a power-law distribution in momentum. This means that a gradial
flattening of the energy distribution is expected at  particle energies $E\lesssim m_{p,e} c^2$.

The prediction of DSA of  power-law distribution in energy with slope $q=2$---for $E>m_{\rm p,e}c^2$---matches
well with the average radio spectral index of $\alpha=0.5$ for
SNRs, given that  for synchrotron radiation the relation between
the particle index for the electron energy distribution and
radio spectral index is $\alpha=(q-1)/2$. Moreover,
the spectral index for cosmic rays in the galaxy is steeper
than the source spectrum due to the energy dependent diffusion
coefficient for cosmic rays, leading to energy dependent escape
of cosmic rays from the galaxy. The relation is $q_{\rm cr}=q_{\rm source}+\delta$, with $\delta$ as defined above.

\subsection{Acceleration timescales and maximum energies}
\label{sec:spatial_time}
It can be shown that for steady state, plane parallel shock the particle distribution
as a function of space coordinate $z$ is
\begin{align}\label{eq:spatial_dist}
    n(z,p) = &n_0(p)\exp[-|z|/l_{\rm diff,1}(p)]~\mbox{for}\ z>0\\
    n(z,p) = &n_0(p)~\mbox{for}~z<0,
\end{align}
with the shock moving in the positive z-direction, and
with $l_{\rm diff}(p)=D_1(p)/v_1=D_1(p)/V_{\rm s}$.
Here $D_1(p)=\frac{1}{3}\lambda_{\rm mfp}v_{\rm cr}$ the diffusion coefficient
in the unshocked (upstream) medium. Here $v_{\rm cr}$ is the particle speed
(for simplicity we take it to be $c$ and assume particle are relativistic), and $\lambda_{\rm mfp}$ the typical
scattering length scale of the particles (mean free path). 
This shows that there are always accelerated particles in front of the shock, with a
length scale that depends on the momentum/energy of the particles.
These particles are referred to as the {\em cosmic-ray shock precursor}.
\index{cosmic-ray precursor}

The shortest scattering length scale possible is of the order of the gyroradius of the particle
$\lambda_{\rm mfp}\gtrsim pc/eB$, and is often parameterised as
$\lambda_{\rm mfp}= \eta pc/eB\approx \eta E/eZB$, with $\eta=1$ referred to
as Bohm diffusion ($Z$ is the charge).
Also downstream there is a typical length scale over which particles are still
able to diffusive back to the shock front: $l_{\rm dif,2}(p)=D_2(p)/v_2\approx \eta ceE/v_2eB$.

The timescale in which particles are typically scattered back over the shock front
on either side of the shock is $\tau_{1,2}=l_{\rm diff,1,2}/v_{cr}$.
From this one can derive the time it takes for particle to perform a full acceleration cycle $\tau=\tau_1+\tau_2$, and the 
acceleration rate can  be derived from
\begin{equation}
    \frac{dE}{dt} = \frac{\Delta E}{\tau_1+\tau_2}= 
    \frac{\frac{4}{3}\frac{\Delta v}{c}E}{\frac{4}{c}\left(\frac{D_1}{v_1}+\frac{D_2}{v_2}\right)}\approx 
    \frac{v_1-v_2}{3}\frac{E}{ \left(\frac{D_1}{v_1}+\frac{D_2}{v_2}\right)}.
\end{equation}
The acceleration time to go from an energy $E_0$ to $E$ is then
\begin{equation}
    t_{\rm acc} = \frac{3}{v_1-v_2}\int_{E_0}^{E}
    \left(
    \frac{D_1}{v_1} + \frac{D_2}{v_2}
    \right) \frac{dE^\prime}{E^\prime}.
\end{equation}
The relation between $v_1$ and $v_2$ depends on the compression ratio $\chi$,
whereas the magnetic-field compression ratio, $\chi_{\rm B}$,
depends on the orientation of the
magnetic field, and is one for parallel magnetic fields and $\chi$ for
perpendicular magnetic fields. Ignoring that $\eta$ may also differ at either
side of the shock, but parametrizing  the diffusion coefficient as
\begin{equation}\label{eq:diff}
    D_1=\frac{1}{3}\eta_{\rm max}%
    \left( \frac{E}{E_{\rm max}}
    \right)^{\delta-1}
    \frac{cE}{eZB_1},
\end{equation}
one can derive the following approximation for the acceleration time
scale to some maximum energy $E_{\rm max}$:
\begin{equation}\label{eq:acc_time}
t_{\rm acc} \approx
1014
\frac{\eta_{\rm max}}{Z\delta}
\left(\frac{E_{\rm max}}{10^{14}~{\rm eV}}\right)
\left(\frac{V_{\rm s}}{5000~{\rm km\,s^{-1}}}\right)^{-2}
\left(\frac{B_1}{10~{\rm \mu G}}\right)^{-1}
\frac{\left(1+\frac{\chi}{\chi_{\rm B}}\right)\left(\frac{\chi}{\chi-1}\right)}{8/3}
~{\rm yr}.
\end{equation}
This shows that $E_{\rm max}\propto \eta_{\rm max}^{-1}B_1V_{\rm s}^2t_{\rm acc}$. 
The numerical values also show that it is quite challenging for SNRs
to accelerated particles up to the cosmic-ray``knee" at $3\times 10^{15}$~eV.
First of all, at an age of $\sim 1000$~yr  the shock velocity of most SNRs
will have dropped to below 5000~\kms. One could also question how
realistic it is that $\eta_{\rm max}=1$, which is the minimum
value possible (corresponding to Bohm diffusion). The only way
to allow acceleration up to the``knee" is if the shock speeds can be higher, which happens in the very early phases of SNR evolution, or if 
$B_1\gg 10{\rm \mu G}$, i.e. much higher than the typical interstellar
magnetic-field strength. 
Amplification of the  magnetic fields near SNR shocks is theoretical possible, 
and one often invoked process is the Bell-mechanism \citep{bell04}, in which the
currents driven by cosmic-ray streaming in the upstream magnetic fields leads
to a plasma instability that amplifies the magnetic field.

\subsection{The effects of radiative losses and cosmic-ray escape on the maximum energy}
\label{sec:escape_losses}
\index{cosmic-ray escape}

According to (\ref{eq:acc_time}) the maximum energy particle can be accelerated to is limited by time. However, there are several other constraining elements.
First of all, electrons radiate quite efficiently due to synchrotron radiation and inverse Compton scattering (Sect.~\ref{sec:radiation}).
We can estimate the maximum energy of electrons in the presence of radiative losses
by equating (\ref{eq:acc_time}) with the synchrotron radiation loss time (\ref{eq:syn_losses}), which gives
\begin{equation}
    E_{\rm e,max}\approx 42 \eta^{-1}\left( \frac{B_2}{100~{\rm \mu G}}\right)^{-1/2}\left(\frac{V_{\rm s}}{5000~{\rm km\,s^{-1}}}\right) TeV,
\end{equation}
where we used now the downstream magnetic-field $B_2=\chi_{\rm B}\chi_{\rm B}$.
Interestingly, converting this maximum energy to the typical synchrotron photon
energy gives a value that no longer depends on the magnetic-field strength \citep{aharonian99}:
\begin{equation}\label{eq:syncutoff}
    h\nu_{\rm max} \approx 3\eta^{-1}
    \left(\frac{V_{\rm s}}{5000~{\rm km~s^{-1}}}\right)^2~{\rm keV}.
\end{equation}
Synchrotron spectra for which the cutoff energy is determined by radiative 
losses are called loss-limited spectra, whereas if it is determined by
the available acceleration time the spectra are called age-limited spectra. \index{age-limited spectra}
\index{loss-limited spectra}
In general, it is now expected that most historical SNRs have loss-limited X-ray synchrotron spectra. But reality is more complex, as shown in a study
of the X-ray synchrotron spectra from Tycho's SNR \citep{lopez15}.

Apart from available time and radiative losses, the maximum acceleration energy
can also be limited by escape, i.e. at very high energies the particles are
no longer expected to return back from the upstream region to the shock.

We can estimate the upstream escape by first calculating the maximum size of the cosmic-ray precursor length scale \citep{vinkbook}, from the fact that
$t_{\rm snr}>t_{\rm acc}$:
\begin{equation}
    t_{\rm snr}>t_{\rm acc}>8 \frac{D_1}{V_{\rm s}}^2 =8 \frac{l_{\rm diff,1}}{V_{\rm s}}.
\end{equation}
The factor 8 incorporates the effects of magnetic-field compression at the shock in a conservative sense. Since we can generically approximate $V_{\rm s}=m R_{\rm s}/t_{\rm snr}$, with $m=1$ indicating free-expansion and $m=2/5$ the Sedov-Taylor expansion, we find that  the precursor length scale is constraint to
\begin{equation}\label{eq:ldiff_max}
    l_{\rm diff,1}< \frac{m}{8}R_{\rm s}< 12.5\% R_{\rm s}.
\end{equation}
We can also turn this around: as soon as the precursor length scale $l_{\rm diff,1}=D_1/V_{\rm s}$
becomes larger than the limit in (\ref{eq:ldiff_max}) the highest energy
particles start escaping.  Equation (\ref{eq:diff}) shows that this may happen if the
upstream magnetic-field strength is declining, but also if the diffusion length scale becomes
larger as function of time due to the deceleration of the shock with time.

The equations on maximum energy discussed sofar assume that the shock velocity is more or
less constant, which is fine as long as the acceleration time is short with respect
to the age of the SNR. However, treating the shock velocity dynamically one can
also derive a maximum energy independent on escape \citep[see][for a derivation]{vinkbook}
\begin{equation}
    E_{\rm max} \approx 0.03 \eta^{-1} \frac{eB_1}{c}m\frac{R_0^2}{t_0}\left(\frac{t}{t_0}\right)^{2m-1},
\end{equation}
where the shock radius as a function of age is approximated by $R=R_0(t/t_0)^m$. 
What is interesting here is that the for $m>1/2$  the maximum energy will always increase
with time for a constant or increasing $B_1$, whereas for $m<1/2$ the maximum energy
is in fact decreasing with time.

\subsection{Non-linear cosmic-ray acceleration}
\label{sec:EfficientAcc}
\index{non-linear diffusive shock acceleration}
In the standard theory of DSA, usually referred to as the test-particle approach, the
shock structure is determined by the standard Rankine-Hugoniot shock equations.
However, if a substantial fraction of the available shock energy flux ($\frac{1}{2}\rho_0 V_{\rm s}^3$) is used to accelerate particles, the shock structure itself will change,
which in turn will change the particle acceleration properties \citep{eichler79}. The theory describing this is non-linear diffusive shock
acceleration (non-linear DSA).

The change itself comes about by the substantial pressure imposed by
the shock precursor (Sect.~\ref{sec:spatial_time}) 
on the plasma upstream of the
shock, due to the fact that the cosmic-ray particles scatter off the
magnetic-field irregularities thereby on average pushing the plasma
away from the shock. This results in the plasma being forced to move before the shock arrives, and is accompanied by a pre-compression of the plasma. This is a consequence of particle-flux conservation, which implies $\rho_0 v_0=\rho_1 v_1$, with $\rho_0$, $v_0$ the plasma density and velocity (in the frame of the shock) 
far upstream of the precursor, and $\rho_1,v_1$ 
the same quantities somewhere in the precursor. The length scale of the precursor depends on the energy of the particles under consideration (\ref{eq:spatial_dist}), as $l_{\rm diff,1}$ is energy-dependent: within
the precursor the cosmic-ray pressure increases toward the shock, and hence
also the precursor compression builds up. The implication is that
as the actual shock arrives (often called the {\em subshock}) the
plasma is already moving with $|v_1|=|v_0/\chi_1<|V_{\rm s}|$, 
with $\chi_1=\rho_1/\rho_0=v_0/v_1$ 
the compression ratio in the precursor. 
Also the gas pressure
is higher than far upstream due to adiabatic compression and perhaps
additional heating mechanisms ($P_1> P_0 \chi_1^{-\gamma}$). Since
the sonic Mach number is given by $M^2=\rho v^2/P$, we have
$M_1^2=\rho_1 v_1^2/P_1 < \rho_0 v_0^2 \chi_1^{-\gamma+1}/P_0 = M_0^2 \chi_1^{-\gamma+1}$, with $\gamma$ the gas-adiabatic index.
The lower Mach number at the (sub)shock or gas-shock results in a lower
plasma temperature. 

Another consequence is that the  DSA process itself needs to be modified: the lowest energy cosmic rays do not diffuse far into
the precursor and sample a lower contrast in shock difference across
the shock ($\Delta v= |v_1 -v_2|$), whereas only the highest energy particles sample the full velocity difference ($\Delta v=|v_0-v_2|$).
The result is that the momentum distribution  of accelerated particles
is no longer a power-law distribution, as indicated by (\ref{eq:dsa_index}), 
but is concave \citep[e.g.][]{ellison00,blasi05}. The total compression
ratio of the shock ($\chi_{\rm tot}=\chi_1\chi_2$) 
can in principle be much larger than the canonical value of 4, and the
asymptotic value of $q=1.5$ \citep{malkov01} 
rather than the canonical $q=2$.

Over the last decade, however, it has become clear that the effects
of non-linear DSA are not as extreme as once predicted. What appears
to hold it in check is another cosmic-ray feedback effect \citep[e.g.][]{vladimirov08,caprioli14}: magnetic-field amplification. This prevents
too much of the precursor compression.

\subsection{The injection problem}
The theory of DSA assumes a population of particles that are elastically
scattering and moving fast enough to be able to scatter from the shocked
medium, back upstream to the unshocked medium (or the precursor).
The theory DSA does not explain how these particles acquired enough energy to start participating in the
DSA process, i.e.
it assumes there is a population of suprathermal particles which form the seed population of DSA accelerated particles.
Early ideas on this "injection problem" assumed that the seed particles were just the high-velocity outliers from a thermal (Maxwellian) distribution
($N(E)\propto E^{1/2}\exp(-E/kT)$), with $E\gg 3 kT$.
However, SNR shocks are collisionless shocks, and it is not a priori
clear whether the shock-heated particles even have a Maxwellian distribution.
A more recent idea about the injection focusses on the phenomenology
of collisionless shocks as seen in particle-in-cell (PIC) calculations,
and in situe measurements of interplanetary  shocks. These show that
instead of immediately shock-heating the incoming particles, collisionless
shocks reflect a fraction of the incoming particles back upstream,
which interact with the unshocked particles. Some of these reflected
particles may cross the shock and be reflected again. According to
PIC simulations \citep{caprioli15} these particles are the ones
that form the seed particles for DSA.

These seed particles are of observational interest as well, as
suprathermal particles may have an effect on the H$\alpha$ line width
for non-radiative, or Balmer-dominated, shocks \citep{nikolic13}, and they may give rise
to measurable suprathermal bremsstrahlung \citep{asvarov90}. In fact,
this was once taken as an alternative explanation for hard X-ray emission
from young SNRs \citep{asvarov90,favata97}, and it is also thought to
be the caused of hard X-ray emission from W49B \citep{tanaka18}.
Note, however, that due to Coulomb collisions the initial spectrum
of the seed particles may be substantially altered further downstream \citep{vink08a}. Hard X-ray emission from these seed particles is clearly
a matter that needs further study, as it provides a glimpse into the initial
stages of cosmic-ray acceleration by SNRs.

\section{X-ray and gamma-ray evidence for cosmic-ray acceleration}
\subsection{Radio and X-ray synchrotron}
\label{sec:xray}

Traditionally SNRs are identified from other shell-type objects, such as HII regions, through
their polarized radio emission and steep spectra in the $10^7$--$10^{11}$~Hz band. These characteristics identify the emission as being caused by synchrotron radiation from nonthermal electrons with energies of around a few GeV. See Eq.(\ref{eq.sync-spec}).
See \cite{dubner15} for a review on radio observations of SNRs.

    Since typical spectral index $\alpha$ is $\sim 0.5$ \cite{green19}, the particle distribution index of accelerated electrons $\Gamma$ can be estimated as $\sim 2$, which is consistent with the DSA theory.
    
    The first discovery of synchrotron X-rays from shells of SNRs was found in shells of SN~1006 \citep{koyama95}, which provided the first evidence of accelerated electrons up to energies of several TeV.
    Currently around ten young SNRs with synchrotron X-rays are listed.
    Interestingly, many of these SNRs have faint or no thermal X-rays
    \cite{koyama97,tsunemi99,bamba00,bamba01,yamaguchi04,reynolds08b,bamba12}, implying that the ambient density of these SNRs are lower than ordninal SNRs with thermal X-rays. 
    The reason may be that X-ray synchrotron radiation requires a high shock speed (Eq.~\ref{eq:syncutoff}), which is more easily maintained over durations of a thousand year for low ambient densities.

    The steepening of the synchrotron spectrum
    corresponding to the cutoff energy of the nonthermal electron distribution, is often occurring in  the soft X-ray band \cite[for example]{bamba08}.
    The cut-off photon energy is proportional to square of the maximum electron energy and magnetic field strength
    (see eq.(\ref{eq.sync-spec})), and measuring the cut-off leads us understanding these physical parameters.
    Assuming the magnetic field of $\sim 10~\mu$G, the maximum energy of electrons is derived to be $\sim 10$--100~TeV, which is one or two orders below the``knee"  energy. 
    In the case of a loss-limited spectrum the photon energy corresponding to the cutoff energy is independent of the magnetic field, and only
    depends on the turbulence parameter $\eta$ and shock velocity, see Eq.~\ref{eq:syncutoff}. Since for several young SNRs the shock velocity has been measured one can infer from the detection of X-ray synchrotron radiation that the magnetic field turbulence must be high, $\eta\lesssim 5$, suggesting that the diffusion of 10--100~TeV electrons is close to the Bohm limit (Sect.~\ref{sec:spatial_time}).
    
    \subsection{X-ray synchrotron radiation from the vicinity of reverse shocks}
   In general it is assumed that the synchrotron emission originates from electron populations accelerated at the forward shock. The reason is that the forward shock velocity is a high Mach number shock for a relatively
   long time ($\sim 10^4$~yr), and that the plasma entering the shock should have typical magnetic-field strength of the order of the average Galactic magnetic field ($B\sim 5~{\rm \mu G}$). However, there is clear evidence
   that part of the plasma immediately downstream of the reverse shock of Cas A is emitting X-ray synchrotron radiation \citep{helder08,uchiyama08}. The implication is also that the bright radio shell of this remnant has electron populations likely accelerated by the reverse shock. Cas A is rather unique in this sense, perhaps because the reverse shock has an inward motion on the western side of Cas A \citep{sato18,vink22a}, which
   makes for a head-on collision between the inward motion of the reverse shock and expanding unshocked ejecta. 
   However, for a few other SNRs synchrotron emission from the reverse shock has also been suggested, such as
   G1.8+0.9 \citep{brose19}, RCW 86/SN 185 \citep{rho02}, and possibly RX J1713.7-3946 \citep{zirakashvili10}.
   
    \subsection{GeV-TeV gamma-rays}
        \label{sec:gevtev}
    
    \begin{figure}
        \centering
        \includegraphics[width=0.6\textwidth]{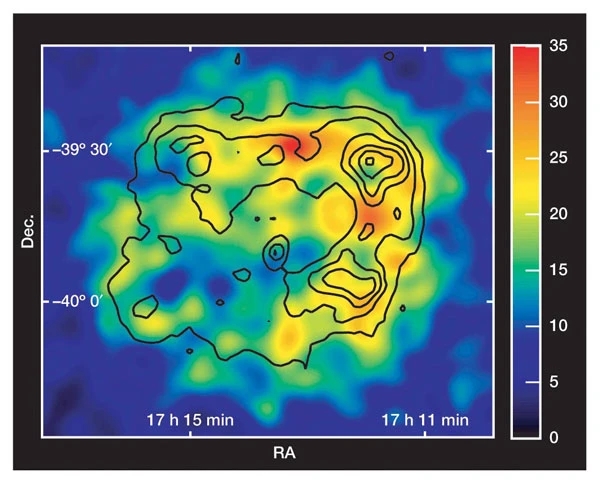}
        \caption{TeV image of the supernova remnant RX~J1713$-$3946 taken by H.E.S.S. \cite{aharonian04}.}
        \label{fig:rxj1713}
    \end{figure}
     \index{H.E.S.S. (High-energy Stereoscopic System}
        
     The first firm detection of gamma-rays from the shells of SNRs was in 2004 by the atmospheric Cherenkov gamma-ray telescope H.E.S.S. \cite{aharonian04} in the very-high-energy (VHE) gamma-ray band (Fig.~\ref{fig:rxj1713}).
     Later, {\it Fermi} detected several SNRs in the high-energy (HE)---or GeV--- band. \index{Fermi satellite}
     Currently more than 30 SNRs are detected in the GeV and/or VHE gamma-ray bands \cite{acero16}.
    In the HE and VHE bands, there are mainly two emission mechanism from accelerated particles, inverse Compton scattering and $\pi^0$ decay emission (Sect.~\ref{sec:radiation}).
   Distinguishing between the leptonic  (inverse Compton) and hadronic scenarios ($\pi^0$ decay) is mainly done with HE band spectra, since a simple leptonic scenario 
     reproduces hard gamma-ray emission with photon index $\Gamma\approx 1.6$ as described in \S~\ref{sec:radiation}, whereas the hadronic scenario requires rather flat spectra ($\Gamma\approx 2.2$) with the low-energy cut off around 100~MeV (the pion bump, \S~\ref{sec:pions}).
     
\begin{figure}
         \centering
         \includegraphics[width=0.7\textwidth]{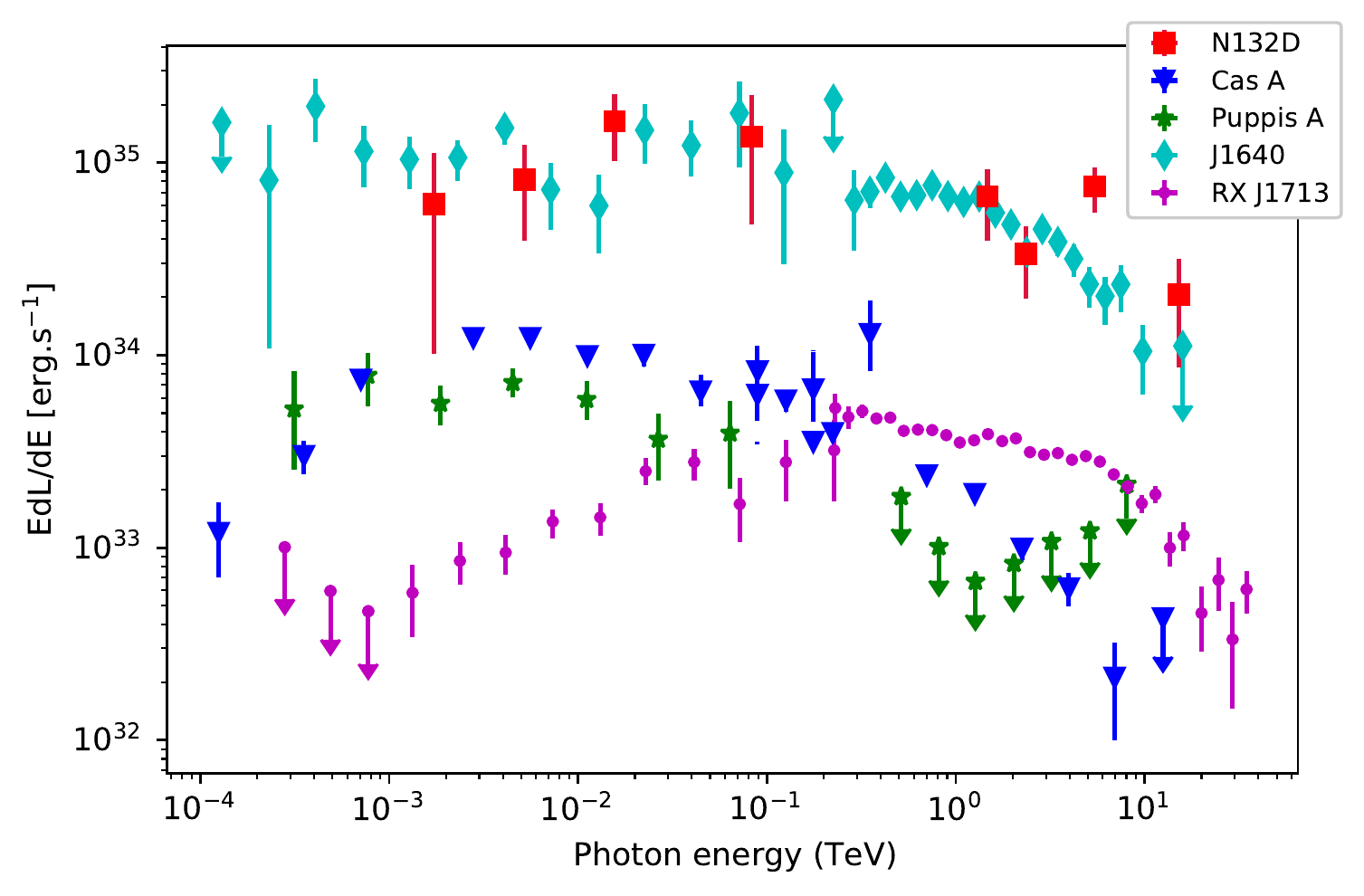}
         \caption{
         Spectral energy diagram (SED) showing the spectral characteristics of various SNRs that are bright in gamma rays.  Most of these SNRs have a spectral index around $\Gamma\approx 2$ below 0.1~TeV, but an exception is RX J1713$-$3946.
         The relatively old SNR Puppis A shows a cutoff around 0.1~TeV, but the other SNRs have spectra steepening above $\sim 1$~TeV
         (Figure reproduced from \citep{hess_n132d}.)
         }
         \label{fig:tev_seds}
     \end{figure}

     A typical case for the leptonic scenario is SN~1006 \cite{acero10}, and for the hadronic scenario W44 provide a good example \cite{abdo10a,giuliani11}.
     On the other hand, there are both leptonic and hadronic components in many SNR cases, and it makes difficult to resolve them. 
     Molecular or atomic cloud observations are another key, since we can measure the density and total amount of target protons. RX~J1713$-$3946, due its gamma-ray brightness a key target to study particle acceleration, shows similar a morphology in VHE gamma rays and proton column density, which is a strong evidence of hadronic origin \cite{fukui12}.
     However, its rather hard gamma-ray spectrum---$\Gamma\approx 1.5$ \citep{fermirxj1713,rxj1713hess18}---suggests a leptonic origin \citep{ellison12}, although more complex multi-zone hadronic models, which include
     dense clumps,  can fit the spectrum as well \citep{inoue12,gabici14}.
     See Fig.~\ref{fig:tev_seds} for a comparison of several gamma-ray luminous SNRs. 
     
     The future Cherenkov Telescope Array (CTA)
     will provide
     VHE gamma-ray maps of fainter SNRs with better spatial resolution and sensitivity. Comparing them with those in synchrotron X-rays (as the electron tracer) and proton column density, more detailed study will be available on leptonic and hadronic components \cite{acero17b}. In addition, we may  be able to observe haloes around SNRs from cosmic rays that have escape the vicinity of the SNR shocks.

    \subsection{Measurements of the cosmic-ray acceleration efficiency}

The total fraction of the supernova explosion energy  that is transferred to cosmic rays is still an open issue. To explain the Galactic cosmic ray energy budget, 5\%--10\% of the canonical $10^{51}$~erg  explosion energy has to be channeled to hadronic cosmic rays.

\begin{figure}
   \centering
       \includegraphics[width=0.6\textwidth]{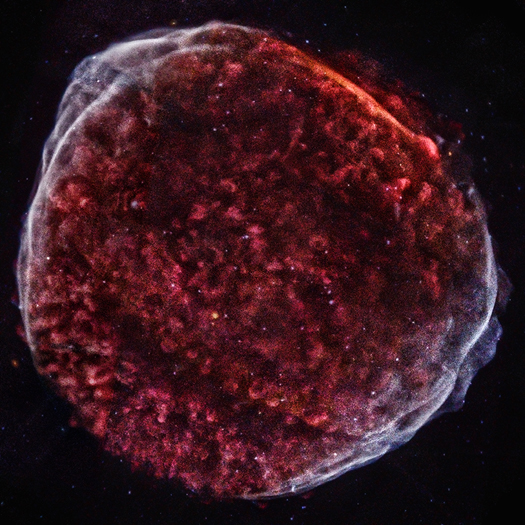}
       \caption{{\it Chandra} view of SN1006. White and thin rim represents synchrotron X-rays, whereas red inside thermal plasma.
       credit: NASA/CXC/Middlebury College/F.Winkler}
       \label{fig:sn1006}
   \end{figure}

 The first evidence for efficient {\em electron} acceleration comes from
 Chandra X-ray observations.
 Chandra discovered that synchrotron X-ray emission is confined to 
 thin filaments---$10^{17}$--$5\times 10^{17}$~cm----near the shocks of young SNRs, such as Cas~A and SN1006 as shown in Fig.~\ref{fig:sn1006} \citep{vink03a,bamba03,bamba05}.
 Thin filaments imply that electrons are trapped near the shock fronts during a radiative loss timescale,
 due to strong and turbulent magnetic fields.
  The strong  magnetic-field turbulence is probably a result of the interaction of  cosmic rays with the plasma in the cosmic-ray shock precursor, as described in \S\ref{sec:EfficientAcc}.
 So the magnetic fields are both an ingredient for, and a result of efficient cosmic-ray acceleration.
 
 Chandra also discovered that
 some synchrotron filaments and knots are 
 time variable with the timescale of $\sim$1~yr
 \citep{uchiyama07,uchiyama08,okuno20,matsuda20}.
 The flux decline implies fast synchrotron losses,
 which also implies fast 
 acceleration, in order for the electrons to reach these high energy in the presence of strong radiative losses.
 The expected magnetic-field strengths are up to a few $100~\mu$G, although there is still many uncertainty on the topology of the magnetic fields.

   Efficient (nonlinear) acceleration results in deviation of the particle spectrum from a pure power law (\S~\ref{sec:EfficientAcc}), which should be reflected in the
   the wide-band synchrotron spectrum.
    Vink et al. \cite{vink06a} measured a spectral bending of the broad, radio-to-X-ray, synchrotron spectrum  of     the young SNR RCW~86, which implies that the acceleration  in this SNR is efficient.
    Similar effects have been reported for SN1006 and Cas A \citep{allen08,domcek21}, which suggests that the efficient acceleration of electrons is common in young SNRs.

The efficiency of hadronic cosmic-ray 
acceleration  cannot be measured directly with X-rays. 
As discussed in \S\ref{sec:pions}, protons interacting with the interstellar medium emit $\sim$GeV gamma-rays via pion decay.
Thus the luminosity in the gamma-ray band reflects
the total energy of accelerated protons, although inverse Compton emission from accelerated electrons contaminates in some cases.
Typical luminosities in the gamma-ray band are
roughly $10^{33}$--$10^{36}$~erg~s$^{-1}$ \citep{acero16},
and the expected proton energy content is $\sim 10^{49}$--$10^{50}$~ergs,   1\%--10\% of the $10^{51}$~erg explosion energy.
One of the most luminous SNRs in the GeV and TeV band is N~132D
in the Large Magellanic Cloud \citep{ackermann16,hess15},
implying N~132D is one of the most efficient proton accelerators.
Given the absence of synchrotron X-ray emission \citep{bamba18} and the broad band spectral shape,
the GeV--TeV gamma-ray emission is best explained
with a hadronic gamma-ray model, with an inferred energy in cosmic-ray protons of 
$\sim 10^{50}$~erg,
which is 10\% of the canonical SN explosion energy
\citep{bamba18,hess_n132d}.
This implies that N~132D is a efficient accelerator 
and/or the explosion energy is larger than $10^{51}$~erg. 

    \subsection{Evidence or lack of evidence for PeVatrons}

Up to now
there is no evidence that some SNRs are PeVatrons.
The current maximum energy of accelerated particles can be estimated from the cut-off of gamma-ray emission.
Considering particle escape, younger SNRs should have higher cut-off.
Acero et al. (2016) \cite{acero16} showed that younger SNRs have harder emission in the GeV band (Figure~16 in \cite{acero16}), although the scatter is very large.
One of the youngest ($\sim 2000$~yrs old) and brighest gamma-ray SNRs, RX~J1713.7$-$3946,
has the cut-off of 3--17~TeV depending on the cut-off models
\cite{hess18_1713}.
This leads the cut-off of proton energy of less than 100~TeV,
indicating that RX~J1713.7$-$3946 does not contain PeV protons, but it may have contained them in the past.
The TeV gamma-ray emission extends significantly beyond the synchrotron X-ray emitting shell,
implying that protons start escaping from the acceleration site, whereas electrons emitting synchrotron X-ray cannot escape due to the synchrotron cooling.
Thus we need even younger samples.
Cas~A is the best target with $\sim$400~years old, 
which is also bright in gamma-rays. 
The spectral break is $\sim$0.2--3~TeV depending on the observations and spectral models \cite{ahnen2017,abeysekara2020}
--- implying that Cas~A does not contribute
cosmic rays in the PeV range.

Even younger samples, G1.9+0.3 ($\sim$100~years old, \cite{reynolds08b}) or SN~1987A ($\sim$40~years old),
are too distant to detect high energy gamma-rays
with the present very high energy gamma-ray telescopes.
Future observations may conclude this problem
with these objects.

Interestingly, 
the aforementioned efficient accelerator, N~132D, shows no evidence for a break in its VHE gamma-ray spectrum---at least up to 8~TeV \citep{hess_n132d}---implying a proton cut-off energy above 45~TeV---see the SED of N132D in Fig.~\ref{fig:tev_seds}.
This is quite remarkable as N~132D is much older (2500~yr) than Cas A, which has a cut-off below 3~TeV.

It is, therefore, likely that other than age, there are other factor determining the maximum acceleration energy, and subsequent escape of these highest energy cosmic rays.

It has long been argued that the highest energy Galactic cosmic rays may come from a subset of SNRs, 
those interacting with the dense winds of their progenitors,
and that the maximum energy is reached within a few years after the SN explosion \citep{ptuskin03,cardillo15,marcowith18}.
At the same time, there is evidence based on the cosmic-ray spectrum itself, that protons may already deviate from a power-law distribution well before the cosmic-ray``knee" \citep{lipari20}.
This could well indicate that only a subset of SNe/SNRs are capable of accelerating up to the knee.

Galactic PeVatrons do exist, as evidenced by the detection of PeV photons from  Galactic plane regions by LHAASO \citep{lhaaso21}. 
One of the PeVatrons is the Crab Nebula, likely a leptonic cosmic-ray sources. Other PeVatrons are associated with star forming regions, but the LHAASO angular resolution is too low to identify individual sources within them. Alternatively, collective effects from fast stellar winds and SNe could create an environment sustained acceleration within superbubbles, i.e. the highly energetic, high pressure regions created by the
collective effects of SN explosions and fast stellar winds. Superbubble acceleration provides an alternative model for the origin of PeV cosmic rays \citep{bykov92,parizot04}.

    \subsection{Evidence for low-energy cosmic rays}
    
    Synchrotron, inverse Compton, and pion decay emissions origin from particles accelerated well into the relativistic regime. On the other hand, there is only a little observation evidence for  the low-energy  particles that have just been  injected into the DSA process.  Nevertheless,
    these low energy cosmic rays may be important for the overall nonthermal energy budget of SNRs.
    
    Low energy protons with energies around $\sim$MeV interact with neutral iron in the interstellar medium and result in characteristic X-ray line emission at 6.4~keV \citep{lodders03,dogiel11,tatischeff12}. Thus 6.4~keV line emission should be a good indicator of $\sim$MeV protons, although the expected surface brightness is very low.
    XIS onboard {\it Suzaku} has a low and stable background level, and discovered the 6.4~keV line from shock-cloud interacting regions of several SNRs \citep{nobukawa15,nobukawa18,bamba18}, which is the first observational clue for sub-relativistic protons.
    The proton energy density is estimated to be $\ge$10--100~eV~cm$^{-3}$, which is more than 10 times higher than that in the ambient interstellar medium.
    
    Electrons just injected have supra-thermal energyies. In the case that their density is enough high, they emit nonthermal bremsstrahlung in the hard X-ray band (Sect.~\ref{sec:bremss}).
    W49B is the only target from which nonthermal bremsstrahlung from suprathermal electrons have been found \citep{tanaka18},
    which is the first clue of electrons that have just been injected into the DSA process.
    More samples with better statistics will be needed 
    to compare to cosmic-ray injection theories.

    \subsection{Cosmic-ray escape from acceleration sites}
    
    Accelerated particles should escape from the shock front in order to become Galactic cosmic rays. Recent observations revealed  some clues for particle escape.
    
    Accelerated electrons are trapped in  the shock region, thus synchrotron X-rays are a good indicator of the shock front. H.E.S.S. found that the VHE gamma rays in one region of RX~J1713$-$3946 are also emitted from a region significantly outside the shock, as marked synchrotron X-rays \cite{hess18_1713}, implying that the detected cosmic rays have already started to escape from the shock.
    
    GeV spectra of middle aged SNRs are rather soft with the cut-off of around 10~GeV \cite{abdo10a},
    which is much smaller than the``knee" energy. This is because that higher energy particles are already escaped from the acceleration site. 
    The cut-off energy becomes smaller as SNRs ages older \cite{acero16,zeng19,suzuki20}. This means that particles with higher energies escape the shock earlier and low-energy particles still remain close to the acceleration sites.
    
    Among the bright GeV SNRs is the mixed-morphology SNR W28. This SNR
    is also detected in VHE gamma rays, but one of the surprises is that two
    VHE sources are not coincident with the SNR, but with nearby molecular clouds \citep{hess08_w28}, which suggests that these clouds are illuminated by hadronic cosmic rays which escaped the SNR.
    
    Particles after the escape diffuse into the interstellar medium. When they encounter dense material, they emit gamma rays via $\pi^0$ decay. Such gamma-ray emitting clouds are found in the vicinity of several SNRs \cite{abdo10b,uchiyama12}.

    \subsection{Polarimetry and magnetic-field turbulence and topology}

As described in \S~\ref{sec:escape_losses}, the emission of X-ray synchrotron by young SNRs requires a turbulent magnetic field in the unshocked medium, see Eq.~\ref{eq:syncutoff}. 
Indeed, radio polarimetric observations of young SNRs show a low polarization fraction of 5--10\% \citep{dickel90}. Another characteristic of young
SNRs, as revealed by radio polarimetry, is that the magnetic-field topology shows a preferentially radial
direction of the magnetic field, whereas older SNRs show a preferentially  tangential magnetic-field orientation \citep{milne75}.
The latter can be easily understood, as shock compression of the magnetic field enhances the
tangential component. 

The radial magnetic-field orientation is not well
understood, but most theories invoke radial stretching of magnetic fields due to hydrodynamical
instabilities such as Rayleigh-Taylor instabilities at the SNR contact discontinuity \citep{gull73,jun96}, or Richtmeyer-Meshkov instabilities \citep{inoue13} near the forward shock.
The former should result in radial magnetic field further away from the shock front.
Alternatively, the radial magnetic field orientation may be an artefact of the acceleration process, if electrons are preferentially accelerated wherever the magnetic field is parallel to the shock normal, biasing the polarization measurements to certain pockets of plasma with a radial magnetic-field direction \citep{west17}.

Up to now these polarimetric measurements were confined to radio synchrotron radiation, and a few cases in the infrared \citep[e.g.][]{jones03}. 
With the launch of the NASA/ASI Imaging X-ray Polarimetric Explorer \citep[IXPE][]{ixpe22} the magnetic-field topology in young SNRs can also be explored in X-rays. This is of interest because,
as we explained above, X-ray synchrotron emission
comes from a region very close to the shock front.
Moreover, magnetic-field turbulence on a long
length scale could leave pockets of plasma
with a particular polarization mode, and a high (up to 40\%) polarization fraction \citep{bykov20}.
However, the presence of these high polarized regions depends on the spectrum of magnetic-field turbulence,
and whether close to the shock the compression imparted a preferred magnetic-field orientation.

At the time of this writing observations of only one SNR, Cas A, have been reported \citep{vink22b}. Surprisingly, it shows a polarization fraction as low, or even lower, as in the radio ($\lesssim 4\%$),
and an overall radial magnetic field structure.
The latter implies that whatever is responsible for the overall radially oriented magnetic-field structure in Cas A, must already have created this pattern within the $\sim 10^{17}$~cm wide X-ray synchrotron rims.

Planned, future observations by IXPE of other SNRs will tell whether Cas A is peculiar, or whether these findings also apply to Tycho's SNR and SN 1006.

\section{Concluding remarks}

In this chapter we introduced the basic X-ray and gamma-ray radiation processes related to the nonthermal particle distirbutions (i.e. cosmic rays) in SNRs.
SNRs are the most plausibly the dominant sources of Galactic cosmic rays, both from an energy budget point of view, but, as we showed, SNRs also show ample
observational evidence for being sites of cosmic-ray acceleration, based on radio, X-ray and gamma-ray observations. As we have shown, 
only gamma-ray radiation can be used to directly probe the hadronic cosmic-ray populations inside SNRs. On the other hand, X-ray synchrotron emission from electrons (leptonic cosmic rays)
provides unequivocal proof that SNR shocks are the sites of active cosmic-ray acceleration.

Despite the overwhelming evidence for cosmic-ray acceleration by SNRs, there is no evidence yet that SNRs are accelerating cosmic rays up to , or beyond, the cosmic ray ``knee”
at $3\times 10^{15}$~eV. In other words, SNRs do not seem to be PeVatrons.
The highest energy cosmic rays leave  SNRs while SNRs are still relatively young ($\lesssim 2000$~yr), as is also evidence by gamma-ray observations. 
But even young SNRs contain cosmic ray particles only with energies below $\sim 100$~TeV. 
The escape of the highest energy particles at a young age suggests that perhaps the highest energy Galactic cosmic rays originate from the very early stages of SNR development,
from a subset of SNRs. On the other hand, competing theories are that stellar superbubbles, or pulsars are responsible for the acceleration of particles up to or beyond the ``knee”.
SNRs are still an important, or even dominant,  contributor to the energization of superbubbles, but the highly turbulent plasma in superbubbles, confining cosmic rays for millions of years,
may be a site for further acceleration of seed cosmic ray particles originating from SNRs and colliding stellar winds.

On the other hand, recent studies of the cosmic-ray spectrum at Earth itself suggest that the composition and spectral shape of the cosmic-ray spectrum around the ``knee” is  more complex
than was assumed until recently, leaving room for cosmic-ray contributions from a mix of SNRs and other (related?) sources.

A number of new X-ray and gamma-ray observing facilities will likely shed further light on the cosmic-ray acceleration properties of SNRs, their cosmic-ray energy budgets, and the role of SNRs
in acceleration particles up to the knee. The recently launched {\it IXPE} satellite is carrying out a program to measure the polarization signatures of X-ray synchrotron radiation, which will inform us
on the magnetic-field turbulence and magnetic-field topology near SNR shocks, both of which are important ingredients for the DSA process.
{\it XRISM} \cite{tashiro2020} (to be launched in 2023) and {\it Athena} \cite{barret2018} (to be launched beyond 2032) will employ X-ray micro-calorimeters with high spectral resolution,
which will enable us to precisely  measure the plasma conditions (electron and ion temperatures)  in the downstream regions of shock, which will lead to a better understanding of
the energy fraction transferred to accelerated particles instead of to the heating of the plasma.
\index{XRISM}
\index{Athena}

Future gamma-ray facilities such as the next generation imaging atmospheric telecope array CTA \index{CTA}
\index{Cherenkov Telescope Array (CTA)}
\cite{2019scta.book.....C} (operational sometime around 2027), and the water Cherenkov telescope LHAASO \cite{2016NPPP..279..166D}, will be able to make deeper probes of the gamma-ray source populations in the Milky Way and the Magellanic Clouds.
LHAASO, which is already partially completed, is in particular very sensitive above 100~TeV, is excellent suited to find Galactic PeVatrons. The first detections \citep{lhaaso21} show the association of the highest energy gamma-ray photons with pulsars and star forming regions. It is still being debated whether the pulsars are purely leptonic cosmic-ray sources, or also sites hadronic acceleration. As for the starforming regions,
here LHAASO lacks the angular resolution to determine whether the region as a whole is a PeVatron---in accordance with the superbubble theory about the origin of cosmic rays---or whether there are PeVatrons
located in these regions, which could include SNRs.
CTA is a pointed observatory, and will be sensitive to photon energies of $\sim$200~TeV, but with an angular resolution of 1--3\arcmin\ above 1~TeV. So it will be excellently suited to pinpoint the sources of
hadronic cosmic rays, even in crowded starforming regions. Moreover, its large effective area is also well suited for transient gamma-ray sources, 
which can be used to search for gamma-ray emission from nearby extragalactic supernovae up to a year after explosion \citep[c.f.][]{hess_supernovae19}.




\printindex

\end{document}